\documentclass[11pt]{article}

\usepackage{
graphicx,amsmath,amssymb,bm}

\newtheorem{prop}{Prop}[section]
\newtheorem{lemma}[prop]{Lemma}

\newtheorem{theorem}[prop]{Theorem}

\title{Universality of replica-symmetry breaking in the transverse field  Sherrington-Kirkpatrick  model
}
\author{C. Itoi, H.  Ishimori, K. Sato and Y. Sakamoto
} 
\begin{document}
\maketitle


\maketitle
\begin{abstract}{ The  existence theorem for replica-symmetry breaking (RSB)  in the  transverse field
 Sherrington-Kirkpatrick (SK) model  is  extended  to the model with a general  random exchange interactions. 
 The relation between the expectation value of the exchange interaction energy  and 
 the Duhamel  correlation function  of spin operators  can be  obtained by  an approximate  integration by parts for general random interactions.
 In addition to the Falk-Bruch inequality,  these explicit evaluations
 enable us to prove that  the variance of overlap between two replica spin operators
 does not vanish
 under sufficiently weak transverse field in sufficiently low temperature.   
The absence of the ferromagnetic long range order is also shown
to distinguish  RSB from  the $\mathbb Z_2$-symmetry breaking.  }
\end{abstract}
\section{Introduction}
Replica-symmetry breaking (RSB)  is studied  extensively as  a 
spontaneous breaking of the replica-symmetry in spin systems with random interactions.
This phenomenon  has been studied  deeply  in mean field classical spin glass models, since Talagrand proved  the Parisi formula \cite{Pr} for the Sherrington-Kirkpatrick (SK) model \cite{SK}  rigorously \cite{T2}. 
It is known that the replica-symmetry breaking phase includes the spin glass phase and a part of the ferromagnetic phase  in the SK model. 
Quite recently,  Leschke, Manai, Ruder and  Warzel  have  proven a remarkable theorem that replica-symmetry breaking phase exists in the transverse field SK model with  centered Gaussian random interactions \cite{W}.  It is shown that the variance of the longitudinal spin overlap does not vanish
 in the system under sufficiently weak transverse field in sufficiently low temperatures. 
 This is a first rigorous result for replica-symmetry breaking in quantum disordered systems.
 
In the present paper, we extend their theorem to non-Gaussian random exchange interactions satisfying an arbitrary symmetric distribution.
It is pointed out that the ferromagnetic long range order gives a non-zero variance of longitudinal spin overlap. 
This paper is organized as follows: 
In section 2,  the Hamiltonian and  physical quantities are defined  and the main theorem for RSB existence 
in the transverse field SK model is described.
In section 3,  the main theorem is proven by several lemmas.  
Finally, several possible extension of the main theorem is discussed.

\section{Definitions and main result}

We study disordered quantum spin systems. A sequence of spin operators  $(\sigma^{w}_i)_{w=x,y,z, i = 1,2, \cdots, N}$
on a Hilbert space ${\cal H} :=\bigotimes_{i =1}^N {\cal H}_i$ is
defined by a tensor product of the Pauli matrix $\sigma^w$ acting on ${\cal H}_i \simeq {\mathbb C}^{2}$ and unities.
These operators are self-adjoint and satisfy the commutation relations
$$
 [\sigma_k^y,\sigma_j^z]=2i \delta_{k,j} \sigma_j^x ,\ \ \  \ \ 
[\sigma_k^z,\sigma_j^x]=2i \delta_{k,j} \sigma_j^y ,\ \  \ \ \ [\sigma_k^x,\sigma_j^y]=2i \delta_{k,j} \sigma_j^z ,  
$$
and each spin operator satisfies
$$
(\sigma_j^w)^2 = {\bf 1}.
$$
The following Hamiltonian with  coupling constants
$h, J \in {\mathbb R}$
\begin{equation}
H_N(\sigma, h, J, \gamma):= J U_N(\sigma^z,\gamma) 
- h \sum_{j=1}^N  \sigma_j^x, 
\label{hamil}
\end{equation}
consists of exchange interactions $U_N$ defined by
\begin{eqnarray}
U_N(\sigma^z, \gamma) :=- \frac{ 1}{\sqrt{N}} \sum_{1\leq i<j\leq N}  \gamma_{i,j}  \sigma_i^z \sigma_j^z
\label{WN}
\end{eqnarray}
where   ${\bm \gamma}:=(\gamma_{i,j})_{1\leq i<j\leq N}$ is a sequence of 
independent identically distributed random variables 
satisfying a probability density function
\begin{equation}
P(\bm \gamma)=\prod_{1\leq i<j\leq N} p(\gamma_{i,j}).
\label{distribution}
\end{equation}
Here, $\mathbb E$ denotes sample expectation of a function $f(\bm \gamma)$ over the sequence $\bm \gamma$  
$$
\mathbb E f(\bm \gamma):= \int d\bm \gamma P(\bm \gamma) f(\bm \gamma).
$$
 Assume that the probability function $p(\gamma_{i,j})$ 
 of each $\gamma_{i,j}\ (1 \leq i<j \leq N)$  is even  function and 
 each moment is given by
 $$
 \mathbb E \gamma_{i,j} =0,  \  \mathbb E \gamma_{i,j}^2 = 1, 
  \ \mathbb E |\gamma_{i,j}|^3 < \infty.
 $$ 
Note that the Hamiltonian  is invariant under $\mathbb Z_2$-symmetry $U \sigma_i ^z U^\dag = -\sigma_i^z$ for
the discrete unitary transformation $U:= \exp \Big( i \pi/2 \sum_{i=1}^N \sigma_i^x\Big) $. 

Here, we define Gibbs state for the Hamiltonian.
For a positive $\beta $,  the  partition function is defined by
\begin{equation}
Z_N(\beta, h, J, \bm \gamma) := {\rm Tr} e^{ - \beta H_N(\sigma,h, J, \bm  \gamma)}\end{equation}
where the trace is taken over the Hilbert space ${\cal H}$.


Let  $f$ be an arbitrary function 
of a sequence of spin operators $\sigma=(\sigma_i^w)_{ w=x,y,z, i=1,2,3, \cdots, N}$.  
The  expectation of $f$ in the Gibbs state is given by
\begin{equation}
\langle f(\sigma) \rangle=\frac{1}{Z_N(\beta, h, J,\bm \gamma)}{\rm Tr} f(\sigma)  e^{ - \beta H_N(\sigma, h, J, \bm \gamma)}.
\end{equation}
Here, we introduce a fictitious time  $t \in [0,1]$ and define a time evolution of operators with the Hamiltonian.
Let $O$ be an arbitrary linear  operator on the Hilbert space $\cal H$, and we define an operator valued function  $O(t)$ of $t\in[0,1]$  by
\begin{equation}
  O(t):= e^{-\beta tH_N} O  e^{\beta tH_N}.
\end{equation}
 The  Duhamel function 
  by 
$$
( O_1,  O_2,\cdots,   O_k) :=\int_{[0, 1]^k} dt_1\cdots dt_k \langle {\rm T}[ O_1(t_1)  O_2(t_2) \cdots  O_k(t_k) ]\rangle,
$$
where 
$ O_1(t_1),  \cdots,  O_k(t_k)$  are  time dependent operators, and 
the symbol ${\rm T}$ is a multilinear mapping of the chronological ordering.
If we define a partition function with arbitrary linear operators  $O_0, O_1, \cdots, O_k$ on the Hilbert space $\cal H$ and real
numbers $x_1, \cdots, x_k$
$$
Z(x_1,\cdots, x_k) := {\rm Tr} \exp \beta \left[O_0+\sum_{i=1} ^k x_i O_i \right],
$$
the Duhamel function of $k$ operators represents
 the $k$-th order derivative of the partition function 
 \cite{Cr,GUW,S}
$$\beta^k( O_1,\cdots,  O_k)=\frac{1}{Z}
\frac{\partial ^k Z}{\partial x_1 \cdots \partial  x_k}.
$$
Also, the connected Duhamel function is defined by
$$\beta^k( O_1; \cdots ;  O_k)=
\frac{\partial ^k }{\partial x_1 \cdots \partial  x_k} \log Z.
$$

To study the spontaneous  $\mathbb Z_2$-symmetry breaking,  define order operator 
\begin{equation}
m= \frac{1}{N} \sum_{i=1}^N \sigma_i^z.
\end{equation} 
Note that  
$\langle \sigma_i ^z \rangle =0$ and thus  $\langle m \rangle=0$ in 
the $\mathbb Z_2$-symmetric Gibbs state for any temperature.

To study replica-symmetry breaking, we consider $n$ replicated spin model defined by the following
Hamiltonian
\begin{equation}
\sum_{a=1} ^n H_N(\sigma^a,h,J, \bm \gamma).
\label{hamilrep}
\end{equation}
The overlap operator $R_{a,b}$ between different replicated spins  is defined by
$$R_{a,b}=\frac{1}{N} \sum_{i=1} ^N \sigma^{z, a}_i \sigma^{z,b}_i,$$
for $a,b = 1,2 \cdots, n,$  and $a \neq b$.

{\theorem \label{MT} 
Consider  the transverse field SK model defined by the Hamiltonian
(\ref{hamil}) and its replicated model (\ref{hamilrep}).
In  the infinite volume limit, 
 the variance of the overlap operator calculated in the replica symmetric and ${\mathbb Z}_2$-symmetric
 Gibbs state does not vanish
\begin{equation}
 \liminf_{N \rightarrow \infty }{\mathbb E}\langle ( R_{1,2} -{\mathbb E} \langle R_{1,2} \rangle)^2 \rangle>0,
\end{equation} 
for sufficiently weak coupling constant  $h$ in sufficiently low temperature. 

In addition, the expectation value of the ferromagnetic order operator  vanishes, and
there is no ferromagnetic  long range order  
\begin{equation}
 {\mathbb E }\langle m \rangle=0=\lim_{N\to\infty}{\mathbb E }\langle m^2 \rangle=\lim_{N\to\infty}{\mathbb E }\langle m^4 \rangle,
\label{noLRO}\end{equation}
in the entire region of coupling constant space.
 }\\

Theorem \ref{MT} shows that the overlap operator $R_{1,2}$ is not self-averaging in the  replica symmetric and ${\mathbb Z}_2$-symmetric 
Gibbs state which has no ferromagnetic long range order. 
The Falk-Bruch inequality \cite{FB,R} 
enables us to prove Theorem \ref{MT}.

\section{Proof}

The following lemma proven by
Carmona and Hu \cite{CH} is useful to study the model with random interactions satisfying a general distribution. See also Ref. 
\cite{Chen} by Chen. 
{\lemma \label{AIP} (An approximate integration by parts)\\
Let  $f:\mathbb R \to \mathbb R$ be a function 
with a bounded continuous third-order partial derivative.  Define
the following difference between two  expectation values 
over a random variable $\gamma$   obeying $p(\gamma)$
\begin{equation}
\Delta(f) :=\mathbb E \gamma f(\gamma) - \mathbb E f'( \gamma) .
\end{equation}
The  absolute value of $\Delta(f)$ is bounded by 
\begin{equation}
|\Delta(f) |  \leq    \frac{3}{2}\mathbb E | \gamma |^3 \sup_x| f''(x)|.
\end{equation}
\noindent
Proof. 
}
In the Taylor series of $f(\gamma)$ and $f'(\gamma)$, for any $\gamma \in \mathbb R$, there exist $c, c' \in (0,1)$, such that
\begin{eqnarray}
&&f(\gamma)=f(0)+ \gamma f'(0)+\frac{\gamma^2}{2} f''(c \gamma), \\
&&f'(\gamma)=f'(0)+\gamma f''(c' \gamma),
\end{eqnarray}
$\mathbb E \gamma=0$ and $\mathbb E \gamma^2=1$ give
$$
\Delta (f) = \mathbb E [\gamma f(\gamma) - f'(\gamma)] =
\mathbb E \Big[\frac{\gamma^3}{2} f''(c\gamma) -\gamma f''(c'\gamma) \Big] \leq \mathbb E  \Big( \frac{ |\gamma|^3}{2} +|\gamma| \Big)  \sup_x| f''(x)| \leq   \frac{3}{2}\mathbb E | \gamma |^3 \sup_x| f''(x)|,
$$
since Jensen's inequality gives 
$$\mathbb E |\gamma| =\mathbb E |\gamma|  \mathbb E |\gamma|^2 =\mathbb E (|\gamma|^3)^\frac{1}{3}  
\mathbb E( |\gamma|^3)^\frac{2}{3} \leq  \mathbb E |\gamma| ^3. 
$$
This completes the proof.
$\Box$  \\

First, 
Lemma \ref{sqR} also for 
the model  defined in the Hamiltonian (\ref{hamil}) is proven as in \cite{W}.

{\lemma \label{sqR} There exists a function $\Delta_N$ of the sequence $\bm \gamma$, such that 
the expectation of the square of  spin overlap is represented by 
\begin{equation}
{\mathbb E}\langle  R_{1,2}^2 \rangle
= \frac{N-1}{N}{\mathbb E}
 (A,A) + \frac{2}{\beta JN}  {\mathbb E}
\langle  U_N \rangle -\frac{2}{\beta J} \Delta_N + \frac{1}{N},
\label{RAU}
\end{equation}
where  
\begin{equation}
A:= \sigma_1^z\sigma_2^z,
\label{A}
\end{equation} and
$$
\lim_{N\to\infty} \Delta_N=0.
$$
\noindent
Proof. } The left hand side is represented as  
\begin{eqnarray}
{\mathbb E}\langle  R_{1,2}^2 \rangle &=&\frac{1}{N^2}\sum_{i=1}^N\sum_{j=1}^N{\mathbb E }\langle \sigma_i^{z,1}\sigma_i^{z,2} 
 \sigma_j^{z,1}\sigma_j^{z,2}\rangle
=\frac{1}{N^2}\sum_{i=1}^N\sum_{j=1}^N{\mathbb E }\langle \sigma_i^{z}\sigma_j^{z} \rangle^2
 \nonumber \\
&=& \frac{1}{N} + \frac{N-1}{N}{\mathbb E } \langle \sigma_1^z \sigma_2^z\rangle^2=\frac{1}{N} + \frac{N-1}{N}{\mathbb E}  \langle A\rangle^2.
\label{RA}
\end{eqnarray}
The expectation value of the exchange energy is
\begin{eqnarray}
\frac{1}{N} \mathbb E \langle  U_N \rangle 
&=& -\frac{1}{N^\frac{3}{2}} \sum_{1\leq i<j\leq N}  {\mathbb E} \gamma_{i,j}
\langle  \sigma_i^z \sigma_j^z \rangle \nonumber \\
&=& -\frac{1}{N^\frac{3}{2}} \sum_{1\leq i<j\leq N}\Big[  \frac{\beta J}{\sqrt{N}} {\mathbb E} 
( \sigma_i^z \sigma_j^z ; \sigma_i^z \sigma_j^z ) + \Delta(\langle  \sigma_i^z \sigma_j^z \rangle) \Big] \nonumber \\
&=&  - \frac{\beta J(N-1)}{2N} {\mathbb E} ( \sigma_1^z \sigma_2^z ; \sigma_1^z \sigma_2^z ) 
-\frac{1}{N^\frac{3}{2}} \sum_{1\leq i<j\leq N}\Delta(\langle  \sigma_i^z \sigma_j^z \rangle) \nonumber \\
&=&   - \frac{\beta J(N-1)}{2N} {\mathbb E}[ (A , A ) - \langle A\rangle^2] +\Delta_N,
\label{UA}
\end{eqnarray}
where  
\begin{equation}
\Delta_N := -\frac{1}{N^\frac{3}{2}} \sum_{1\leq i<j\leq N}\Delta(\langle  \sigma_i^z \sigma_j^z \rangle).
\end{equation}
The identity (\ref{RAU}) is obtained   from above 
identities (\ref{RA}) and (\ref{UA}). 
Lemma \ref{AIP} gives the upper bound on $|\Delta ( \langle \sigma_i^z\sigma_j^z \rangle )|$
\begin{eqnarray}
|\Delta ( \langle \sigma_i^z\sigma_j^z \rangle )|
&\leq &   \frac{ 3}{2}  \mathbb E |\gamma_{i,j}|^3
\sup \Big| \frac{\partial^2 }{\partial\gamma_{i,j}^2} \langle \sigma_i^z\sigma_j^z \rangle \Big|  \label{Delta}\\
 &=& \frac{ 3}{2}  \mathbb E |\gamma_{i,j}|^3
  \frac{\beta^2J^2}{N}
  \sup| (\sigma_i^z\sigma_j^z; \sigma_i^z\sigma_j^z;\sigma_i^z\sigma_j^z )|,
 \nonumber
  \\
 &=&
  \frac{3\beta^2J^2c_3}{2N} \mathbb E  |\gamma_{1,2}|^3.
\end{eqnarray}
where  $c_3 := \sup|  (A;A;A)| =\sup| (A,A,A) -3 (A, A) \langle A \rangle + 2 \langle A \rangle^3| \leq 6 $ for $A:= \sigma_1^z \sigma_2^z$. Therefore
\begin{eqnarray}
|\Delta_N| \leq  \frac{1}{N^\frac{3}{2}} \sum_{1\leq i<j\leq N}
  \frac{3\beta^2J^2c_3}{2N}  \mathbb E |\gamma_{i,j}|^3
\leq     \frac{3\beta^2J^2 c_3}{4\sqrt{N}} \Big(1-\frac{1}{N} \Big) \mathbb E  |\gamma_{i,j} |^3
\end{eqnarray} 
Therefore 
$$
\lim_{N\to \infty} \Delta_N =0.
$$
This completes the proof. $\Box$\\

The following lemma is one key which Leschke, Manai, Ruder and Warzel 
have used to prove the non-zero variance of the overlap operator 
\cite{W}. 

{\lemma \label{FB}  The following
 Duhamel function of $A:=\sigma_1^z\sigma_2^z$ is bounded from the below
\begin{equation}
(A,A) \geq \frac{1}{2\beta h}(1-e^{-2\beta h}).
\label{DuhamelA}
\end{equation}

\noindent
Proof.} 
The Falk-Bruch inequality \cite{FB,R} gives  the following lower bound on 
 Duhamel function of the operator $A$ defined by (\ref{A})
\begin{equation}
(A,A) \geq \langle A^2\rangle\Phi\Big( \frac{\beta}{4 \langle A^2\rangle} \langle [ A, [H_N,A]] \rangle \Big),
\end{equation}
where the function $\Phi: [0,\infty) \to [0,\infty) $ is defined by $\Phi( r \tanh r) := \frac{\tanh r}{r}$.
Using  $A^2 = \bm 1$ and 
$[A, [H_N,A]]  =  4 h (\sigma_1^x+ \sigma_2^x)$,   the Falk-Bruch inequality implies
\begin{equation}
(A,A) \geq \Phi (\beta h \langle  (\sigma_1^x+ \sigma_2^x) \rangle ).
\end{equation}
Since the function $\Phi$ is monotonically decreasing and $\langle\sigma_i^x\rangle \leq 1$,
\begin{equation}
(A,A) \geq \Phi (2 \beta h  ).
\label{AA}
\end{equation}
The following lower  bound is given by Dyson, Lieb and Simon \cite{DLS}
$$
\Phi(t) \geq \frac{1}{t} (1-e^{-t}). 
$$
 for $t\geq 0$.  
 This and  the  bound (\ref{AA})  for $(A, A)$ complete the proof. $\Box$\\

The following lemma  is obtained  by Carmona and Hu \cite{CH}  

{\lemma \label{U}
The expectation value of the exchange energy has the  following  lower bound 
 \begin{equation}
\lim_{N\to\infty}\frac{1}{N} {\mathbb E}
\langle  U_N \rangle
\geq  - \kappa,
\label{LBU}
\end{equation}
where  $\kappa  \simeq 0.763 $ is given by the ground state energy
in the SK model with the standard Gaussian r.v.s $\bm g$ \cite{CR,Pr}.   \\

\noindent
Proof. }  
The expectation of exchange energy
is bounded by its  ground state energy
$$
\mathbb E \langle U_N \rangle \geq  \mathbb E 
 \inf_{\langle \sigma | \sigma \rangle=1} \langle \sigma |( - \sum_{i<j} \frac{\gamma_{i,j}}{\sqrt{N}} \sigma_i^z \sigma_j^z) | \sigma \rangle,
$$
where the normalized ground state 
$| \sigma \rangle$  in the SK model
with r.v.s $\bm \gamma$ is given by an eigenstate of each $\sigma_i^z$, such that $\sigma_i^z |\sigma \rangle = \sigma_i |\sigma \rangle$. 
Carmona and Hu give the following identity \cite{CH}
$$
\lim_{N \to \infty} \frac{1}{N}\mathbb E 
 \inf_{\sigma } ( - \sum_{i<j} \frac{\gamma_{i,j}}{\sqrt{N}} \sigma_i \sigma_j) 
 =
 \lim_{N \to \infty} \frac{1}{N}\mathbb E 
 \inf_{\sigma}( - \sum_{i<j} \frac{g_{i,j}}{\sqrt{N}} \sigma_i \sigma_j)  = -\kappa.
 $$
The right hand side is originally given by Parisi \cite{Pr}. 
$\Box$

\paragraph{Proof of Theorem \ref{MT}}
Lemma \ref{sqR}, inequalities (\ref{DuhamelA})  and (\ref{LBU}) imply that
the square of the spin overlap 
 has the  following  expectation  bounded from the below in the infinite volume limit
\begin{equation}
\liminf_{N\to\infty} {\mathbb E}
\langle  R_{1,2}^2 \rangle
\geq  \frac{1}{2 \beta h}(1- e^{-2\beta h}) -\frac{2 \kappa}{\beta J},
\label{LBW}
\end{equation}
where  $\kappa \simeq 0.763$ has been evaluated already \cite{CR,Pr}.   
The $\mathbb Z_2$-symmetry gives  $\langle \sigma_i^z\rangle=0$ and therefore
$$
\langle R_{1,2}\rangle =\frac{1}{N}\sum_{i=1} ^N\ \langle \sigma_i^z \rangle^2 =0.
$$
This and the bound (\ref{LBW}) yield
\begin{equation}
\liminf_{N \rightarrow \infty}
 [  {\mathbb E} \langle {R_{1,2}}^2 \rangle-({\mathbb E} \langle {R_{1,2}}\rangle)^2]>0,
\label{varianceR}
\end{equation}
for sufficiently weak $h$ and sufficiently low temperature.\\
To show no ferromagnetic long range order, let us evaluate the expectation value of $m^2$
\begin{equation}
{\mathbb E }\langle m^2 \rangle = \frac{1}{N^2} \sum_{i=1}^N \sum_{j=1}^N{\mathbb E } \langle \sigma_i^z \sigma_j^z\rangle
=\frac{1}{N} + \frac{N-1}{N}{\mathbb E } \langle \sigma_1^z \sigma_2^z\rangle.
\label{m2}
\end{equation}
For a unitary transformation $U_1:=\exp ( i \frac{\pi}{2} \sigma_1^x)$,  the operator $\sigma_1^z$ is transformed into
$U_1 \sigma_1^zU_1^\dag =-\sigma_1^z$. Since $p(\gamma_{1,j}) = p(-\gamma_{1,j})$ is assumed,  
${\mathbb E } \langle \sigma_1^z \sigma_2^z\rangle={\mathbb E } \langle U_1 \sigma_1^z \sigma_2^z U_1^\dag\rangle=
-{\mathbb E } \langle \sigma_1^z \sigma_2^z\rangle
$
which gives ${\mathbb E } \langle \sigma_1^z \sigma_2^z\rangle=0$ and thus
$$
\lim_{N\to\infty} {\mathbb E }\langle m^2 \rangle =0.
$$
Also, $\mathbb E \langle m^4 \rangle $ can be represented in
\begin{equation}
\mathbb E \langle m^4 \rangle = \frac{N!}{N^4(N-4)!} {\mathbb E} \langle  \sigma_1^z \sigma_2^z \sigma_3^z \sigma_4^z  \rangle
 +  \frac{3N-2}{N^3}.
 \end{equation}
${\mathbb E} \langle  \sigma_1^z \sigma_2^z \sigma_3^z \sigma_4^z  \rangle=0$ is proven 
in the same argument as for ${\mathbb E}\langle   \sigma_1^z \sigma_2^z \rangle =0$. Then,
$$
\mathbb E \langle m^4\rangle=0.
$$ 
 This completes the proof of Theorem \ref{MT}. $\Box$\\

Theorem \ref{MT} implies that  RSB occurs  and 
the variance of $\langle m^2 \rangle $ vanishes
$$
\lim_{N\to\infty} [ \mathbb E \langle m^2 \rangle^2 - (\mathbb E \langle m^2 \rangle)^2]=0,
$$
since $ \langle m^4 \rangle \geq  \langle m^2 \rangle^2.$
 Therefore, the Chebyshev inequality 
 implies that $\langle m^2 \rangle $ vanishes with probability 1 in the infinite volume limit,  and there is no
 ferromagnetic long range order in each sample.

\paragraph{ Discussions}   Here, we discuss several extensions of our result to some other models, where there is a  possibility of
$\mathbb Z_2$-symmetry breaking.

In the model with random exchange interactions satisfying a non-centered distribution $\mathbb E \gamma_{i,j} \neq 0$, 
spontaneous $\mathbb Z_2$-symmetry breaking  can appear.
If  there is ferromagnetic long range order, then ${\mathbb E} \langle \sigma_1^z \sigma_2^z \rangle \neq 0$   gives a finite variance of the overlap
$$
{\mathbb E} \langle R_{1,2}^2 \rangle - ({\mathbb E} \langle R_{1,2} \rangle )^2= \frac{N-1}{N}
{\mathbb E} \langle \sigma_1^z \sigma_2^z \rangle^2  \geq \frac{N-1}{N}( {\mathbb E} \langle \sigma_1^z \sigma_2^z \rangle)^2 \neq 0,
$$
in the $\mathbb Z_2$-symmetric Gibbs state. If  ferromagnetic long range order exists,  
$$
\liminf_{N\to\infty}[{\mathbb E} \langle R_{1,2}^2 \rangle - ({\mathbb E} \langle R_{1,2} \rangle )^2 ]>0,
$$
should be proven in $\mathbb Z_2$-symmetry breaking Gibbs state with 
spontaneous magnetization to show the existence of RSB.  Since    ${\mathbb  E} \langle R_{1,2} \rangle $ 
does not vanish because of the
$\mathbb Z_2$-symmetry breaking $\langle \sigma_i^z \rangle \neq 0$,  the proof becomes nontrivial. 

In models defined by Hamiltonians with $\mathbb Z_2$-symmetry breaking terms,  such as longitudinal fields or 
$p$-spin interactions for an odd positive integer $p$, the order parameter
${\mathbb  E} \langle R_{1,2} \rangle $ does not vanish because of the
$\mathbb Z_2$-symmetry breaking $\langle \sigma_i^z \rangle \neq 0$, then the proof becomes nontrivial also.\\
 

\noindent
{\bf Acknowledgments} We thank the anonymous referee of
 the previous version of the present
paper for 
valuable and constructive comments.
 C.I. is supported by JSPS (21K03393).

\end{document}